# Nonlinear spin and orbital Edelstein effect in WTe$_2$

Xing-Guo Ye,[1] Peng-Fei Zhu,[1] Wen-Zheng Xu,[1] Tong-Yang Zhao,[1] and Zhi-Min Liao[1,2,*]

[1]*State Key Laboratory for Mesoscopic Physics and Frontiers Science Center for Nano-optoelectronics,
School of Physics, Peking University, Beijing 100871, China*
[2]*Hefei National Laboratory, Hefei 230088, China*

In materials with spin-momentum locked spin textures, such as Rashba states and topological surface states, the current-induced shift of the Fermi contour in the $k$ space leads to spin polarization, known as the Edelstein effect, which depends linearly on the applied current. However, its nonlinear counterpart has not yet been discovered. Here, we report the observation of the nonlinear Edelstein effect in few-layer WTe$_2$. Under a current bias, an out-of-plane magnetization is induced in WTe$_2$, which is electrically probed using an Fe$_3$GeTe$_2$ electrode, a van der Waals ferromagnet with perpendicular magnetic anisotropy. Notably, with an applied ac at frequency $\omega$, an induced magnetization with second-harmonic response at frequency $2\omega$ is observed, and its magnitude demonstrates a quadratic dependence on the applied current, characteristic of the nonlinear Edelstein effect. This phenomenon is well explained by the current-induced orbital magnetization via the Berry connection polarizability tensors in WTe$_2$. The orbital degree of freedom plays the primary role in the observed nonlinear Edelstein effect, that is, the nonlinear orbital Edelstein effect. This can, in turn, give rise to a nonlinear spin Edelstein effect through spin-orbit coupling.

The efficient conversion of charge to spin is crucial for developing next-generation spintronic devices [1–12]. This conversion is particularly notable in materials exhibiting spin-momentum locked spin textures, such as those with Rashba spin-orbit coupling (SOC) or topological surface states. In Rashba systems, the breaking of inversion symmetry leads to the splitting of spin bands into inner and outer Fermi circles with opposite spin helicities [5]. Applying a current induces an imbalance in the density of spin-up and spin-down electrons, resulting in net spin polarization. Similarly, topological insulators and topological semimetals, with their unique topological surface states, also exhibit efficient charge-to-spin conversion due to their intrinsic spin-momentum locking [6–8]. This phenomenon, generally known as the Edelstein effect, typically exhibits a linear relationship between the induced spin polarization and the applied current. The linear Edelstein effect has been used to generate spin torques to drive the magnetization switching of ferromagnetic materials, significantly impacting the development of magnetic random-access memory [11].

Despite extensive research on the Edelstein effect, the exploration of its nonlinear counterpart has remained elusive until now. A nonlinear Edelstein effect, where the induced spin polarization depends quadratically on the applied current, would have profound implications. In this study, we present an experimental observation of the nonlinear Edelstein effect in few-layer WTe$_2$. When subjected to an ac bias, WTe$_2$ exhibits an out-of-plane magnetization that displays a second-harmonic response at twice the frequency of the driving current, with a magnitude that scales quadratically with the current—a hallmark of the nonlinear Edelstein effect. This phenomenon is elucidated through the current-induced orbital magnetization due to the Berry connection polarizability (BCP) tensors in WTe$_2$.

*Orbital magnetization and nonlinear Edelstein effect induced by BCP tensors.* Recently, orbital magnetization [13–17], orbital torques [18–20], orbital Hall effect [21–25], and the kinetic magnetoelectric effect [26–33], which are related to the orbital magnetic moment, have garnered significant research interest. Within the framework of semiclassical Boltzmann transport equation along with constant relaxation time approximation, the current-induced orbital magnetization is given by $M_i = e\tau\alpha_{ij}^{\text{ME}}E_j$, where $i$ and $j$ label the spatial directions, $E$ is the electric field, $\tau$ is the scattering time, and $\alpha_{ij}^{\text{ME}}$ is the kinetic magnetoelectric (ME) susceptibility. The $\alpha_{ij}^{\text{ME}} = \int_{\boldsymbol{k}} \frac{df}{d\varepsilon} m_{\boldsymbol{k},i} v_{\boldsymbol{k},j}$ can be notable for an asymmetric orbital texture, where $\frac{df}{d\varepsilon} \propto \delta(\varepsilon - \varepsilon_F)$ is approximately valid at low temperatures $T << T_F$ ($T_F$ is the Fermi temperature), $m_{k,i}$ is the $i$th component of orbital magnetic moment $m_k$, $v_k = \frac{\partial \varepsilon}{\partial k}$ is the electron velocity, and the integral is over the Brillouin zone with summation over the band index [34]. Moreover, the nonlinear magnetoelectric coupling can also lead to current-induced magnetization [35].

In WTe$_2$ with the Td phase, the orbital magnetization can be generated by applying current. Td-phase WTe$_2$ is classified as a type-II Weyl semimetal, showcasing Fermi arc surface states [36]. Remarkably, at the monolayer level, WTe$_2$ exhibits the quantum spin Hall effect [37] and superconductivity [38,39]. The crystal structure of WTe$_2$ belongs to the space group $Pmn2_1$, with mirror symmetry $\mathcal{M}_a$ and glide mirror symmetry $\tilde{\mathcal{M}}_b$ [40]. These two mirror symmetries with perpendicular mirror planes together generate a twofold rotational symmetry $C_{2z}$ in the $ab$ plane through a half-cell translation. By applying a current to break this $C_{2z}$ symmetry, the nonlinear Edelstein effect can be induced through the mechanism of BCP tensors, defined as $G_{ij} = 2e\text{Re}\sum_{m\neq n}\frac{(A_i)_{nm}(A_j)_{mn}}{\varepsilon_n - \varepsilon_m}$, where $(A_i)_{nm} = \langle n|i\frac{\partial}{\partial k_i}|m\rangle$ is the interband Berry connection and $|n\rangle$ is the cell-periodic part of the Bloch eigenstate with eigenenergy $\varepsilon_n$ [41–44]. The BCP tensors are gauge-invariant band geometric quantities associated with the electric-field induced positional shift of Bloch electrons [43]. The BCP is regarded as the intrinsic origin of the third-order nonlinear Hall effect [44]. Upon applying an electric field, the BCP tensor can lead to Berry connection linearly proportional to the applied current, denoted as $\mathbf{A}^{(1)} = \overleftrightarrow{\mathbf{G}}\mathbf{E}$. Here, $\overleftrightarrow{\mathbf{G}}$ is the BCP tensor, $\mathbf{E}$ is the applied electric field, and the superscript "(1)" indicates the first-order term of the electric field. Note that the orbital magnetic moment $\mathbf{m}_k$ involves the matrix elements $-\frac{e}{2}\langle\psi_i|\mathbf{r}\times\mathbf{v}|\psi_i\rangle$, with $\mathbf{r}$ as the position operator, $\mathbf{v} = -\frac{i}{\hbar}[\mathbf{r}, H]$ as the velocity operator, $H$ as the Hamiltonian, and $|\psi_i\rangle$ as the eigenstate [45]. In the Bloch representation, the position operator is formulated as $\mathbf{r} = i\frac{\partial}{\partial \mathbf{k}} + \mathbf{A}(\mathbf{k})$ [46]. Therefore, the BCP tensors can give rise to a field-correction term of $\mathbf{m}_k$ through the field-induced Berry connection $\mathbf{A}^{(1)} = \overleftrightarrow{\mathbf{G}}\mathbf{E}$, that is, $\mathbf{m}_k^{(1)} = -\frac{e}{2}\langle n|(\overleftrightarrow{\mathbf{G}}\mathbf{E})\times\mathbf{v}|n\rangle$, which is linearly proportional to $\mathbf{E}$.

---

*Contact author: liaozm@pku.edu.cn

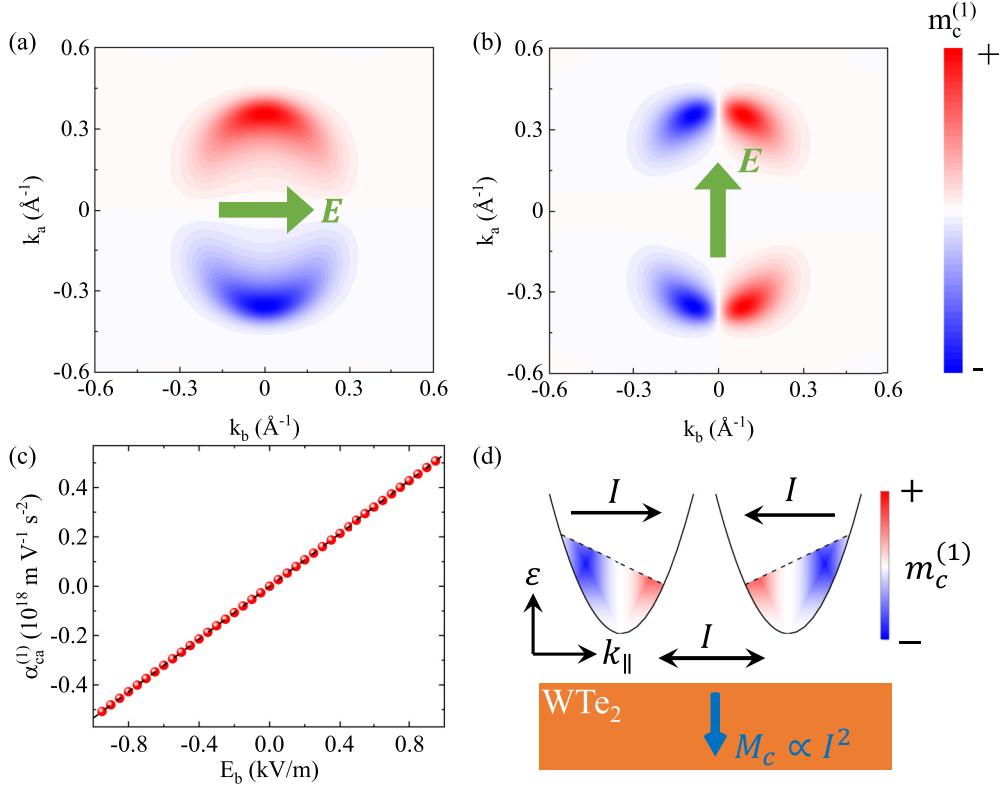

FIG. 1. (a), (b) Illustration of asymmetric distribution of orbital magnetic moment in the conductance band when applying an electric field along (a) $b$ axis or (b) $a$ axis in Td monolayer WTe$_2$, respectively. (c) The coefficient $\alpha_{ca}^{(1)}$ as a function of the in-plane electric field $E_b$ at chemical potential 0.1 eV. (d) Illustration of nonlinear Edelstein effect in WTe$_2$. The applied current induces an orbital texture, further leading to orbital magnetization, which exhibits a quadratic dependence on the current.

Figures 1(a) and 1(b) illustrate the electric-field induced orbital magnetic moments in the Td monolayer WTe$_2$ by applying an electric field along the $b$ axis and $a$ axis, respectively, exhibiting antisymmetric distributions in momentum space, where $m_c^{(1)}(k_b, k_a) = -m_c^{(1)}(k_b, -k_a)$ for $\mathbf{E} = E\hat{\mathbf{b}}$, and $m_c^{(1)}(k_b, k_a) = -m_c^{(1)}(-k_b, k_a)$ for $\mathbf{E} = E\hat{\mathbf{a}}$. This dipolelike distribution forms an exotic orbital texture with orbit-momentum locking [23,32], giving rise to the kinetic magnetoelectric coefficient $\alpha_{cj}^{(1)}$, where an in-plane electric field is able to generate out-of-plane orbital magnetization ($M_c$). Figure 1(c) illustrates the $\alpha_{ca}^{(1)}$ as applying an electric field $\mathbf{E}$ along the $b$ axis in WTe$_2$ (see Supplemental Material, Figs. S1 and S2 [47] for model calculations), giving $\alpha_{ca}^{(1)} \propto E$. Utilizing this orbital texture, the electric field additionally triggers orbital magnetization, denoted as $M_c = e\tau\alpha_{cj}^{(1)}E_j \propto E^2$ [55], constituting a nonlinear Edelstein effect. While the nonlinear orbital Edelstein effect analysis provided relies on the Boltzmann equation method, a more comprehensive theoretical framework, starting from nonlinear transport theory [56], remains desirable. Furthermore, it is noteworthy that with the presence of SOC, the orbital magnetization gives rise to spin polarization, leading to the emergence of the nonlinear spin Edelstein effect. However, it is important to note that the orbital degree of freedom is more fundamental here, as the nonlinear orbital dynamics are independent of the strength of SOC, whereas the nonlinear spin Edelstein effect only appears when SOC is activated [55].

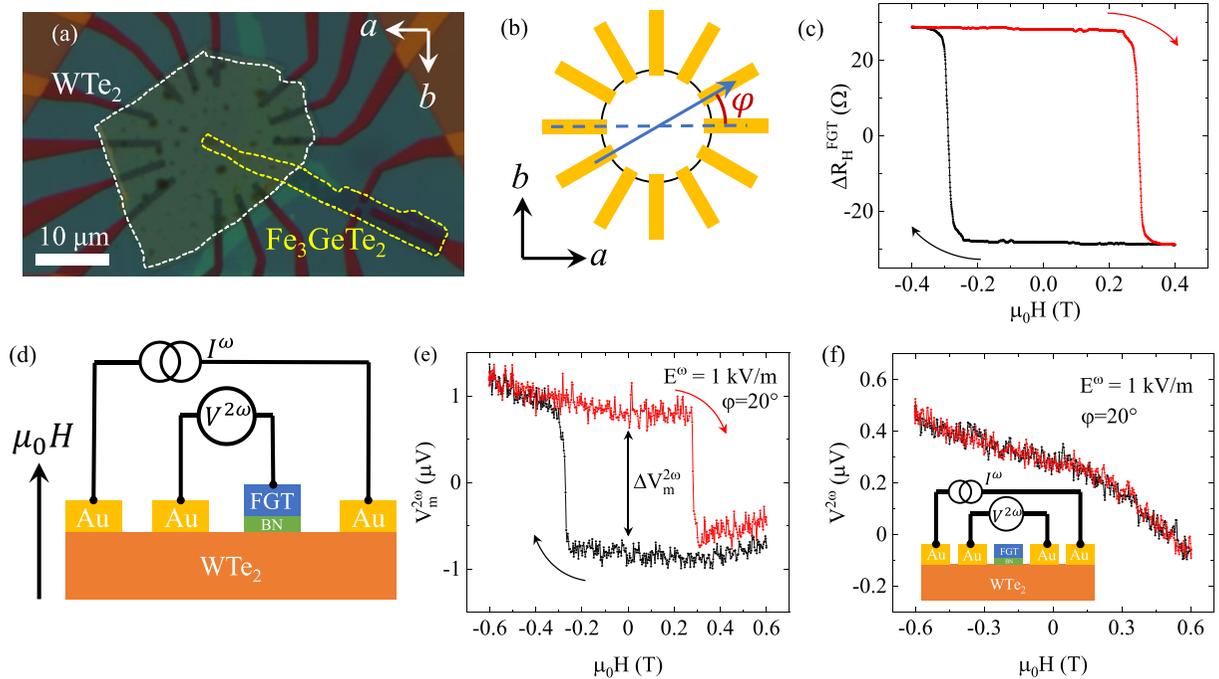

FIG. 2. (a) Optical image of a device. (b) Schematic of circular disc electrodes. The angle $\varphi$ is defined as the angle between the $a$ axis of WTe$_2$ and the direction of applied current. (c) The Hall resistance of FGT, showing hysteresis with magnetic field. (d) Schematic of magnetization detection in WTe$_2$ via FGT probe. (e) Magnetization-related second-harmonic voltage $V_m^{2\omega}$ at $\varphi = 20°$ and $E^\omega = 1$ kV/m. (f) Using two Au electrodes for the voltage measurement at $\varphi = 20°$ and $E^\omega = 1$ kV/m, no hysteresis loop is observed. All data were acquired at 5 K.

The nonlinear Edelstein effect is characterized by $M_i = \chi_{ijk} E_j E_k$ ($i, j, k = a, b, c$ in WTe$_2$, with the $c$ axis as the out-of-plane orientation) [48]. Here, we only consider the tensor component $\chi_{cij}$, specifically for $i, j = a, b$, representing the in-plane field-induced out-of-plane magnetization. The $\chi_{cij}$ is intricately linked to the BCP tensors, allowed by inversion symmetry (refer to the symmetry constraints detailed in Supplemental Material, Table S1 [47]). Since WTe$_2$ bulk holds both $\mathcal{M}_a$ and $\tilde{\mathcal{M}}_b$ mirror symmetries [44], the two mirrors enforce $\chi_{caa} = \chi_{cbb} = 0$, and only the components $\chi_{cab}$ and $\chi_{cba}$ are permissible. Therefore, applying an electric field $E_a$ or $E_b$ along the crystalline axis cannot induce out-of-plane magnetization [44,48]. However, deviating the $E$ from the crystalline $a$- or $b$ axis allows the generation of out-of-plane magnetization $M_c$ due to the nonzero $\chi_{cab}$ and $\chi_{cba}$.

In experiments, the out-of-plane magnetization in WTe$_2$ bulk is generated by applying a charge current $I$ [Fig. 1(d)], effectively generating an electric field denoted as $E = RI/L$, where $R$ represents the longitudinal resistance ($16 \sim 56\,\Omega$ in our device), and $L$ is the channel length ($\sim 8\,\mu$m in our device). The current-induced magnetization is expressed as $M_c = e\tau\alpha_{cj}^{(1)} E_j \propto I^2$, which remains unchanged as reversing the direction of current.

*Current-induced out-of-plane magnetization in* WTe$_2$. To directly measure the current-induced out-of-plane magnetization in WTe$_2$, we employ a few-layer Fe$_3$GeTe$_2$ (FGT) as a ferromagnetic (FM) probe. Results from two devices, namely devices S1 and S2, are presented, with the focus on device S1 in the main text. The optical image of device S1 is depicted in Fig. 2(a). A layer of $h$-BN thin film was interposed between WTe$_2$ and FGT to prevent magnetic proximity effect from FGT. Disc electrodes were fabricated to apply current along various directions, as illustrated in Fig. 2(b), where the angle $\varphi$ represents the relative angle between the applied current and a pair of electrodes (along the crystalline $a$ axis with a small misalignment of $\sim 6°$). The perpendicular magnetism anisotropy of FGT is confirmed by a square-shaped hysteresis loop in the anomalous Hall resistance, as shown in Fig. 2(c) (also refer to Fig. S5 in the Supplemental Material [47]).

The experimental setup for measuring the current-induced magnetization in WTe$_2$ is illustrated in Fig. 2(d). In these measurements, a fixed ac ($I^\omega$) with a frequency $\omega$ is applied to generate magnetization through the nonlinear Edelstein effect. Given that $M_c \propto I^2$, a second-harmonic out-of-plane magnetization $M^{2\omega}$ with a frequency of $2\omega$ is anticipated. The magnetization in WTe$_2$ is then probed using the FGT probe. When $M_c$ is antiparallel or parallel to the magnetization of FGT ($M_{\text{FGT}}$), a high- or low-voltage state will be detected, respectively (see Supplemental Note 3 for detailed mechanisms [47]). By varying $M_{\text{FGT}}$ through sweeping an external magnetic field and simultaneously measuring the voltage $V^{2\omega}$ using the FGT probe with respect to a reference Au electrode via the lock-in method, the current-induced $M_c$ in WTe$_2$ can be directly obtained (see Fig. S6 [47] for the detailed measurement configuration). The magnetization-dependent voltage is determined by subtracting the background from $V^{2\omega}$, denoted as $V_m^{2\omega}$ (refer to Fig. S7 [47]).

The measured $V_m^{2\omega}$ with a current applied at an angle deviated from the $a$ axis ($\varphi = 20°$) is presented in Fig. 2(e). Hysteresis behavior is evident in $V_m^{2\omega}$ during both forward and backward sweeping magnetic fields. The high- and low states of $V_m^{2\omega}$ correspond to the current-induced $M^{2\omega}$ in WTe$_2$ being antiparallel and parallel to $M_{FGT}$, respectively. A difference in $V_m^{2\omega}$ between up- and down sweeping magnetic fields above $+0.3$ T is noted, attributed to the cumulative nature of thermal effects over time under the applied ac field $E^\omega = 1$ kV/m ($\sim 0.5$-mA current). No hysteresis loop is observed for the voltage probe employing two nonmagnetic electrodes with the FGT electrode between them [Fig. 2(f)]. This observation unequivocally dismisses the possibility of the $V_m^{2\omega}$ originating from the proximity-induced magnetism in $WTe_2$ by the FGT electrode.

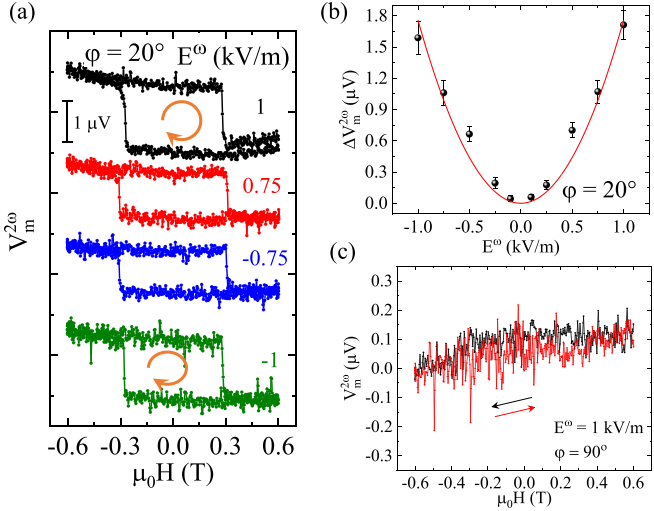

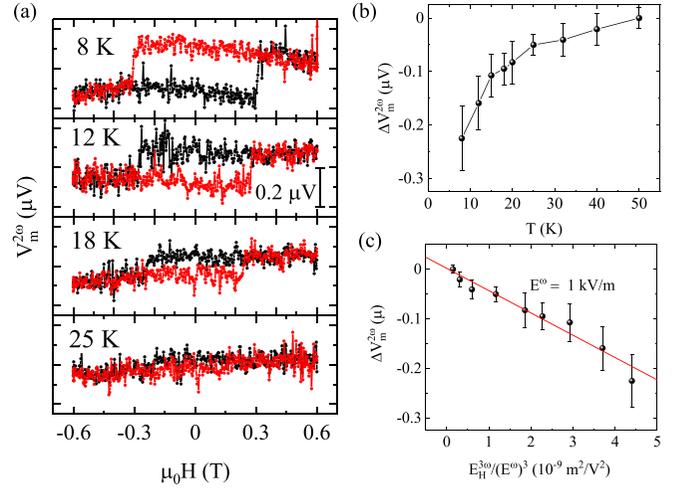

FIG. 3. (a) $V_m^{2\omega}$ under various $E^\omega$ at $\varphi = 20°$. (b) The height of $V_m^{2\omega}$ hysteresis curve, $\Delta V_m^{2\omega}$, as a function of $E^\omega$ at $\varphi = 20°$. The red line is a fit with the formula of $\Delta V_m^{2\omega} = \lambda(E^\omega)^2$. (c) $V_m^{2\omega}$ at $E^\omega = 1$ kV/m with $\varphi = 90°$ (approximately along the $b$ axis). All data were acquired at 5 K.

FIG. 4. (a) Second-harmonic $V_m^{2\omega}$ under $E^\omega = 1$ kV/m along $\varphi = -20°$ at different temperatures as indicated. (b) Temperature dependence of the hysteresis height $\Delta V_m^{2\omega}$. (c) The $\Delta V_m^{2\omega}$ at fixed $E^\omega = 1$ kV/m as a function of the third-order nonlinear Hall signal $\frac{E_H^{3\omega}}{(E^\omega)^3}$ along $\varphi = -20°$. The data were collected across a temperature range of 8 to 50 K.

The $V_m^{2\omega}$ at $\varphi = 20°$ is further investigated by varying the intensity of excitation current, as shown in Fig. 3(a). It is found that the loop height increases with excitation current, and its polarity remains unchanged upon reversing the excitation current (by exchanging the source-drain leads). The loop height $\Delta V_m^{2\omega}$, defined as the voltage difference at zero magnetic field when forward- and backward sweeping magnetic field, exhibits a quadratic dependence on the excitation current [Fig. 3(b)]. Because $\Delta V_m^{2\omega} \propto M_{WTe_2}$, the observations suggest a quadratic relationship between the induced out-of-plane magnetization in $WTe_2$ and the excitation current, aligning with the anticipated behavior of the nonlinear Edelstein effect.

The $V_m^{2\omega}$ is further investigated when applying current $I^\omega$ at $\varphi = 90°$ (approximately along the $b$ axis). As depicted in Fig. 3(c), suppressed hysteresis is observed as sweeping magnetic field. Due to the $bc$ mirror plane imposing $\chi_{cbb} = 0$ in $WTe_2$, inducing out-of-plane magnetization with $I^\omega$ along the $b$ axis is not feasible. The slight, nonvanishing hysteresis in Fig. 3(c) is attributed to a misalignment of $\sim 6°$ between $I^\omega$ and the $b$ axis. Consistent results were obtained with excitation currents along different directions (refer to Fig. S8 [47]) and under various measurement configurations (see Fig. S9 [47]). Furthermore, reproducible results were confirmed in device S2 (refer to Figs. S10 and S11 in Supplemental Material [47]).

The temperature dependence of current-induced magnetization is further investigated. As depicted in Fig. 4, the $V_m^{2\omega}$ diminishes with increasing temperature, and the hysteresis loop becomes barely discernible above 50 K. Concurrently, the third-order nonlinear Hall effect of $WTe_2$ was measured at various temperatures. Remarkably, when plotting $\Delta V_m^{2\omega}$ against $\frac{E_H^{3\omega}}{(E^\omega)^3}$, a clear linear dependence emerges, as illustrated in Fig. 4(c). Here, $E_H^{3\omega} = \frac{V_H^{3\omega}}{W}$, $E^\omega = \frac{I^\omega R_{xx}}{L}$, $V_H^{3\omega}$ is the third-harmonic Hall voltage, and $W$ and $L$ are the channel width and length, respectively. Note that the quantity $\frac{E_H^{3\omega}}{(E^\omega)^3}$ is directly related to the BCP tensors in $WTe_2$ [44]. The BCP tensors contribute to the $\frac{E_H^{3\omega}}{(E^\omega)^3}$ by $\mathcal{X}_{abcd}^1 = 2\int_k f \partial_a \partial_b G_{cd} - \int_k f \partial_c \partial_d G_{ab} - \frac{1}{2}\int_k \frac{d^2 f}{d\varepsilon^2} v_a v_b G_{cd}$, where $G_{ij}$ is the BCP tensors and $\mathcal{X}_{abcd}^1$ is the (intrinsically) $\tau$-independent part of $\frac{E_H^{3\omega}}{(E^\omega)^3}$ [42]. Therefore, the observed linear correlation between $\Delta V_m^{2\omega}$ and $\frac{E_H^{3\omega}}{(E^\omega)^3}$ indicates that the current-induced nonlinear magnetization in $WTe_2$ also possesses an intrinsic origin from the BCP tensors.

*Discussion and conclusions.* The magnetization-dependent second-harmonic voltage signals $\Delta V_m^{2\omega}$ distinctly indicate the nonlinear out-of-plane magnetization induced by in-plane current, that is, the nonlinear Edelstein effect. We also examine other mechanisms contributing to this nonlinear magnetization (details in Supplemental Note 7), and it has been determined that the nonlinear spin- (orbital) Hall effect and spin Seebeck effect are not responsible for the observed phenomena.

Recent reports indicate that in-plane and out-of-plane magnetization with $\Delta V_m \approx 200$ and $30$ μV, respectively, can be induced by the linear Edelstein effect with an applied current of 1 mA along the $a$ axis in $WTe_2$, where the out-of-plane magnetization is allowed by the surface symmetry breaking [51–54,57]. Despite its relatively weaker magnitude observed in this work, the current-induced magnetization from the nonlinear Edelstein effect demonstrates a quadratic relationship with the current, ensuring its direction remains unchanged upon the reversal of the current. Furthermore, the lower-symmetry requirements of the nonlinear Edelstein effect render it more applicable across a broader range of crystalline materials. These characteristics endow the nonlinear Edelstein effect with unique potential applications in charge-to-magnetization conversion.

In summary, our results reveal the emergence of nonlinear Edelstein effect in few-layer WTe$_2$. The application of a current induces an exotic orbital texture through the positional shift of Bloch electrons, further leading to the generation of magnetization. The induced magnetization demonstrates a quadratic dependence on the applied current, directly observed using a ferromagnetic FGT probe with second-harmonic voltage measurements. This finding represents a significant advancement in understanding the nonlinear charge-to-spin conversion mechanisms, particularly highlighting the role of current-induced orbital magnetization facilitated by BCP tensors. Considering the widespread existence of BCP in diverse materials, regardless of SOC [34, 42–44], the nonlinear Edelstein effect holds significant promise in a broad range of materials with weak SOC for the electric control of magnetization.

*Acknowledgments.* This work was supported by the National Natural Science Foundation of China (Grants No. 62425401 and No. 62321004), and Innovation Program for Quantum Science and Technology (Grant No. 2021ZD0302403).

**Supplemental Materials for**

**Nonlinear Spin and Orbital Edelstein Effect in WTe$_2$**

Xing-Guo Ye[1], Peng-Fei Zhu[1], Wen-Zheng Xu[1], Tong-Yang Zhao[1], and Zhi-Min Liao[1,2*]

[1] State Key Laboratory for Mesoscopic Physics and Frontiers Science Center for Nano-optoelectronics, School of Physics, Peking University, Beijing 100871, China.

[2] Hefei National Laboratory, Hefei 230088, China.

*E-mail: liaozm@pku.edu.cn

**Method**

**Device fabrications.** The high-quality bulk WTe$_2$ crystals were obtained from HQ Graphene. Few-layer WTe$_2$ flakes were obtained by a mechanical exfoliation method, where the WTe$_2$ crystal was repeatedly exfoliated with scotch tape, transferred onto polydimethylsiloxane (PDMS), and then covered onto a precleaned Si substrate with 285 nm-thick SiO$_2$. The PDMS was heated for about 1 minute at 90°C to transfer the few-layer WTe$_2$ onto the Si substrate. Ti/Au electrodes (around 10 nm thick) were patterned onto the SiO$_2$/Si substrate using e-beam lithography, metal deposition, and lift-off. The exfoliated h-BN, Fe$_3$GeTe$_2$ (FGT), and few-layer WTe$_2$ were sequentially picked up and transferred onto the Ti/Au electrodes using a polymer-based dry transfer technique.

In the fabrication of the device, ~7-nm-thick WTe$_2$ flake was chosen and was stacked with monolayer or bilayer h-BN and 10-nm-thick FGT strip, where the h-BN served as the spacer to prevent the magnetic proximity effect between the FGT and WTe$_2$. A 20-30 nm thick h-BN layer was placed on top for coverage. The entire exfoliation and transfer process were performed in an argon-filled glove box with an O$_2$ and H$_2$O content below 0.01 parts per million to prevent sample degradation.

**Supplemental Note 1: Berry connection polarizability in Td monolayer WTe2**

The Berry connection polarizability (BCP) tensor is a band geometric quantity that is related to the field-induced positional shift of Bloch electrons, given by $G_{\mu\nu}(n, \boldsymbol{k}) = 2e \text{Re} \sum_{m \neq n} \frac{(A_\mu)_{nm}(A_\nu)_{mn}}{\varepsilon_n - \varepsilon_m}$, where $A_{mn}$ is the interband Berry connection and $e$ is the electron charge [42]. The BCP tensor can generate Berry connection through an external electric field following $\boldsymbol{A}^{(1)} = \overleftrightarrow{\boldsymbol{G}} \boldsymbol{E}$, where $\overleftrightarrow{\boldsymbol{G}}$ is the BCP tensor, $\boldsymbol{E}$ is the external electric field, and the superscript "(1)" represents that the physical quantity is the first order term of electric field.

The BCP tensors are calculated in monolayer Td WTe2 using the effective $\boldsymbol{k} \cdot \boldsymbol{p}$ Hamiltonian

$$\mathcal{H} = \begin{pmatrix} \epsilon_c & v^+ & & 0 \\ -v^- & \epsilon_v & & \\ & & \epsilon_c & v^- \\ 0 & & -v^+ & \epsilon_v \end{pmatrix},$$

where $\epsilon_c = \epsilon_1$, $\epsilon_v = \frac{1}{1+r}\epsilon_2 - \frac{r}{1+r}\epsilon_3$, $v^\pm = \sqrt{\frac{1}{1+r}}(\pm v_{1,x} k_x + i v_{1,y} k_y)$, and $\epsilon_i = c_{i,0} + c_{i,x} k_x^2 + c_{i,y} k_y^2$ ($i = 1,2,3$). All the parameters used are the same as in Ref. [32]. The calculated BCP tensors are shown in Fig. S1. Giant BCP tensors are found near band edge. Importantly, even when the spin-orbit coupling is absent, the BCP tensors are still significant in WTe2.

It is worth noting that the monolayer model is a simplified approach to understanding the electronic properties of the few-layer WTe2. Nevertheless, this model captures the main features of the BCP tensors when compared with the results from first-principles calculations of the bulk system (*ref*. 44). These main features are crucial for understanding the nonlinear magnetization signals observed in our experiments.

However, the monolayer model cannot capture the detailed stacking interactions and variations in electronic properties that arise in the few-layer and bulk systems. First-principles calculations of the bulk system, as shown in *ref.* 44, reveal more detailed features that the monolayer model cannot account for.

The field-induced orbital magnetic moment in monolayer Td WTe$_2$ is calculated by a Kubo-like formula, $\boldsymbol{m}_{\boldsymbol{k}}^{(1)} = -\frac{e}{2\hbar}\langle n|(\overleftrightarrow{\boldsymbol{G}}\boldsymbol{E}) \times \boldsymbol{\nabla}_{\boldsymbol{k}}H|n\rangle$, where the $n$-th band BCP tensors $\overleftrightarrow{\boldsymbol{G}}$ are calculated as $G_{ij} = 2eRe\sum_{m\neq n}\frac{\langle n|\frac{\partial H}{\partial k_i}|m\rangle\langle m|\frac{\partial H}{\partial k_j}|n\rangle}{(\varepsilon_n-\varepsilon_m)^3}$, as shown in Fig. 1 of main text.

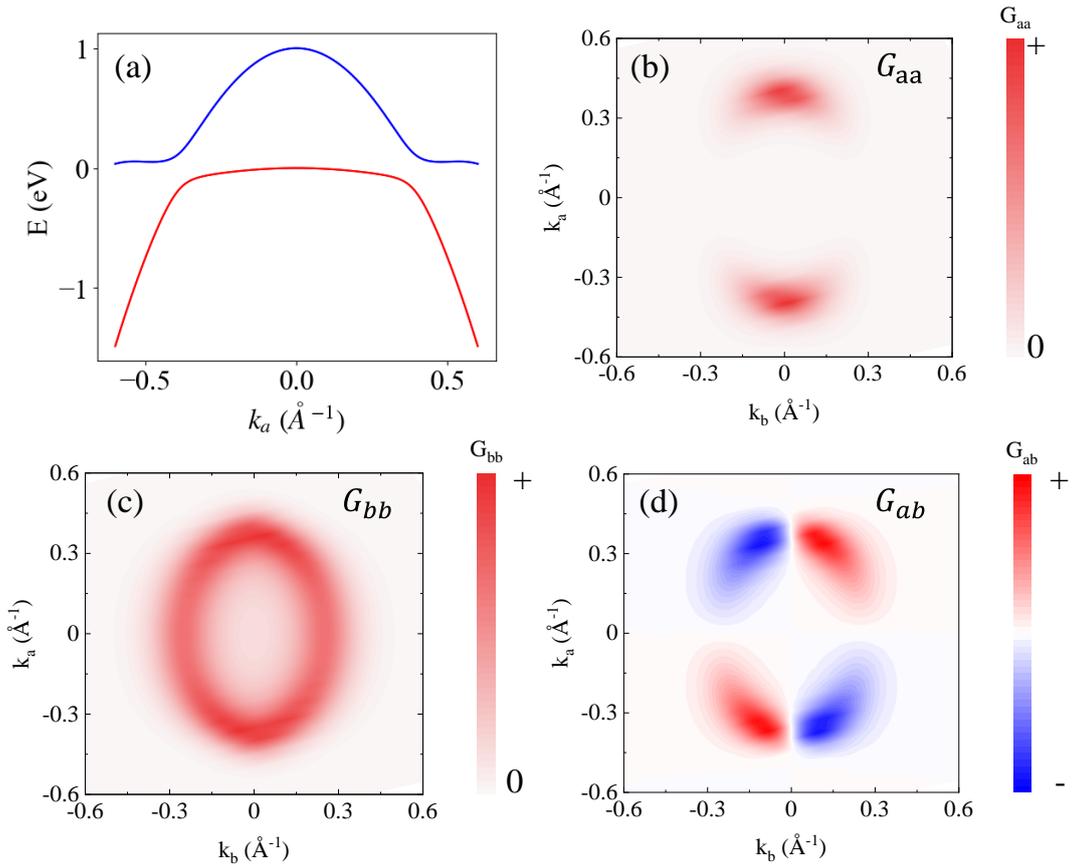

**Fig. S1.** (a) Energy band of the monolayer WTe$_2$, where the spin-orbit coupling is turned off, leading to spin-degenerate bands. (b)-(d) Distribution of the BCP component (b) $G_{aa}$, (c) $G_{bb}$, and (d) $G_{ab}$ in momentum space.

To demonstrate this field-induced orbital magnetic moment, calculations are also implemented in a two-dimensional (2D) gapped Dirac model with the Hamiltonian

$$H(\mathbf{k}) = \xi k_a \sigma_0 + v_a k_a \sigma_a + v_b k_b \sigma_b + \Delta \sigma_c,$$

where $\sigma_0$ is the $2 \times 2$ identity matrix, $\sigma_{a,b,c}$ is Pauli matrices, $\xi$ describes the tilting of Dirac cone, $v_a$ and $v_b$ describe the anisotropy of Fermi velocity, $2\Delta$ is the energy gap [42]. The energy band of the 2D gapped Dirac model is shown in Fig.S2(a). The BCP tensors are evaluated, as shown in Figs. S2(b)-(d). It is found that the BCP tensors show maximum value near the band edge. Under the application of an external electric field, orbital magnetic moment can be generated by the BCP tensors. The field-induced orbital magnetic moment shows a dipole-like distribution in momentum space, leading to nonzero kinetic magnetoelectric coefficient $\alpha_{cj}^{(1)}$, as shown in Figs. S2(e)-(f).

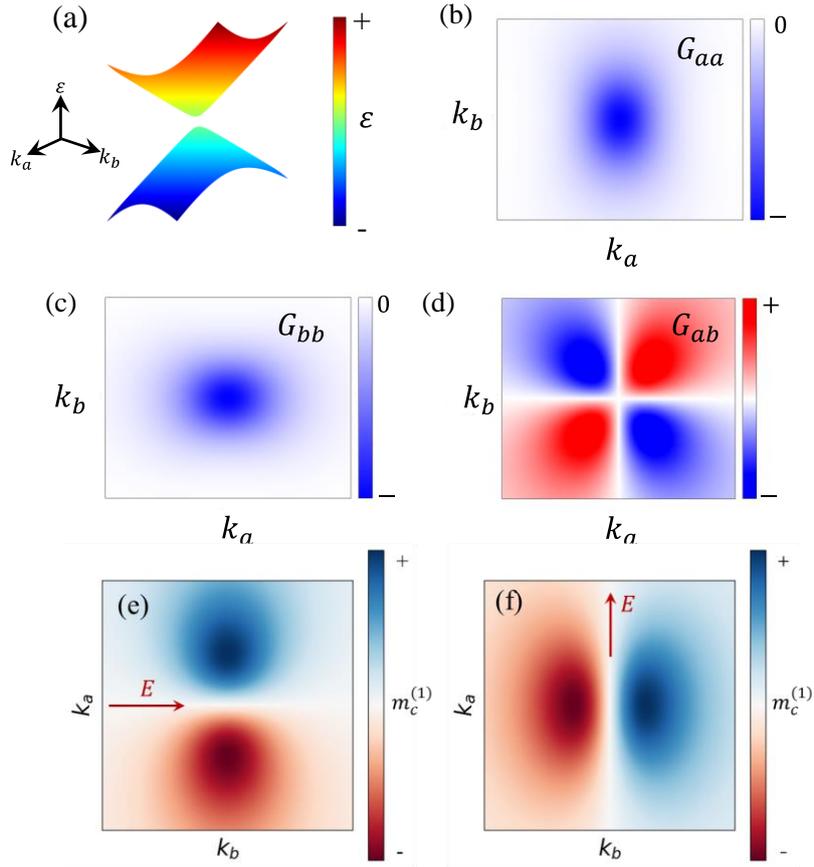

**Fig. S2. The 2D gapped Dirac model.**

**a,** Energy band of the gapped Dirac model.

**b-d,** Distribution of the BCP component (**b**) $G_{aa}$, (**c**) $G_{bb}$, and (**d**) $G_{ab}$ in momentum space.

**e,f,** Field induced orbital magnetic moment under an external electric field applying along the (**e**) *b* axis and (**f**) *a* axis.

## Supplemental Note 2: Nonlinear orbital Edelstein effect

The orbital magnetic moment arises from the self-rotation of Bloch electrons around their center of mass [22,23]. An unusual orbital texture characterized by orbit magnetic moment-momentum locking can result in a nonzero magnetoelectric coefficient $\alpha_{cj}^{ME} = \int_{\bm{k}} \frac{df}{d\varepsilon} m_{\bm{k},c} v_{\bm{k},j}$, and the out-of-plane orbital magnetization is given by $M_c = e\tau \alpha_{cj}^{ME} E_j$ [34], known as the out-of-plane orbital Edelstein effect. However, in materials with inversion symmetry, the $\alpha_{cj}^{ME}$ is forced to be zero, leading to this out-of-plane orbital Edelstein effect vanishing [27]. Alternatively, the $\alpha_{cj}^{ME}$ can be induced by applying an electric field through the mechanism of BCP tensors, where the restriction of inversion symmetry is removed.

As depicted in the upper panel of Fig. S3, the applied current generates an antisymmetric orbital texture with a nonzero $\alpha_{cj}^{(1)} \propto I$ due to the field-correction of Berry connection induced by BCP tensors. The applied current further induces orbital magnetization, as illustrated in the lower panel of Fig. S3, through the Fermi contour shift. Therefore, the current induces orbital magnetization with $M_c = e\tau \alpha_{cj}^{(1)} E_j \propto I^2$, representing the nonlinear orbital Edelstein effect.

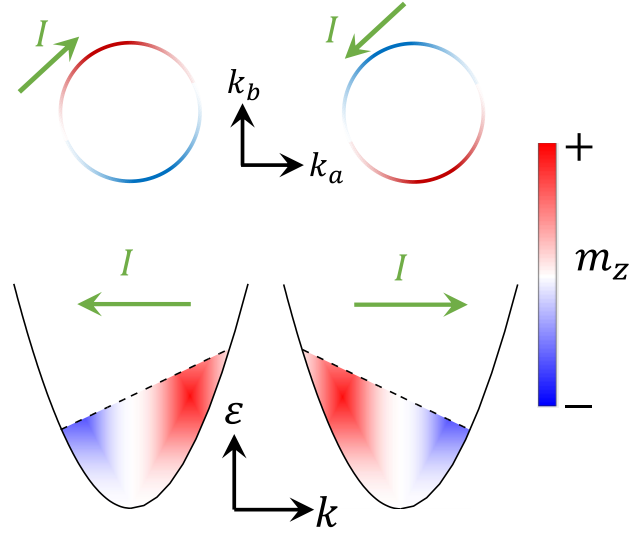

**Fig. S3. Illustration of BCP-induced nonlinear magnetoelectric effect.** Upper panel: The current-generated orbital texture in WTe$_2$ with $\alpha_{cj}^{ME} \propto I$. Lower panel: Illustration of the current-induced orbital magnetization with nonequilibrium electron distribution.

The out-of-plane nonlinear orbital Edelstein effect relies on the mechanism of BCP tensors. Given the widespread presence of BCP tensors across various materials, the nonlinear orbital Edelstein effect is anticipated to manifest in a diverse range of substances. Unlike the linear orbital Edelstein effect exclusive to gyrotropic crystals [27], the nonlinear counterpart exhibits a more lenient symmetry constraint [48]. To provide a generalized framework for the nonlinear orbital Edelstein effect in diverse materials, we summarize the symmetry constraints. Specifically, we consider the time-reversal-even nonlinear magnetoelectric coefficient $\chi_{ijk}$ in nonmagnetic materials, focusing on in-plane electric field-induced out-of-plane magnetization. Assuming the **E** is applied in the $ab$ plane ($j, k = a, b$) and the induced magnetization is along the out-of-plane $c$ axis ($i = c$), the obtained symmetry constraints are presented in Table S1. Notably, Table S1 reveals that the nonlinear magnetoelectric coefficient $\chi_{cab}$ is permissible in WTe$_2$ bulk, precisely aligning with the results obtained in this work.

**TABLE S1.** The symmetry constraints of the nonlinear magnetoelectric coefficient $\chi_{ijk}$ under time-reversal symmetry. Here $\gamma$ is the inversion symmetry, $C_n^i$ is the $n$-fold rotation symmetry with rotation axis along $i$ direction. $\mathcal{M}_i$ is the mirror symmetry with the mirror line along $i$ direction. Allowed: ✔ ; Forbidden: 🚫.

|  | $\gamma, C_2^c$ | $C_{3,4,6}^c$ | $C_{2,4,6}^a$ | $C_3^a$ | $\mathcal{M}_c$ | $\mathcal{M}_a, \mathcal{M}_b$ |
|---|---|---|---|---|---|---|
| $\chi_{caa}$ | ✔ | ✔ | 🚫 | 🚫 | ✔ | 🚫 |
| $\chi_{cbb}$ | ✔ | ✔ | 🚫 | ✔ | ✔ | 🚫 |
| $\chi_{cab}$ | ✔ | 🚫 | ✔ | ✔ | ✔ | ✔ |

In this work, we choose few-layer WTe$_2$ to demonstrate the nonlinear orbital Edelstein effect based on the following considerations:

1. WTe$_2$ exhibits metallic behavior, which facilitates an efficient Fermi contour shift driven by electric current. This shift is crucial for the nonlinear orbital Edelstein effect, representing a kinetic magnetoelectric effect.

2. The bulk of WTe$_2$ has significant BCP tensors, leading to the generation of a Berry curvature dipole under an electric field. This effect has already been considered as the origination of the third-order nonlinear Hall effect observed in WTe$_2$ bulk. Consequently, the nonlinear orbital Edelstein effect can be highlighted as the dominant term.

**Supplemental Note 3: Magnetization detection using a ferromagnetic probe**

The current induced out-of-plane magnetization $M_c$ in WTe₂ is directly measured through the detection of a magnetization-dependent voltage $V_M$ using a ferromagnetic (FM) probe [6-10]. The $V_M$ depends on the relative orientation between the $M_c$ and the magnetization of the FM probe ($M_{FM}$). When $M_c \parallel M_{FM} \parallel -c$, electrons with magnetic moment of $-m_c$ will diffuse from WTe₂ to the FM probe, causing the FM probe to become negatively charged and yielding $V_M < 0$. Conversely, when $M_c \parallel -M_{FM} \parallel -c$, electrons with magnetic moment of $+m_c$ will diffuse from the FM probe to WTe₂, causing the FM probe to become positively charged and yielding $V_M > 0$. These processes are illustrated in Figs. S4(a)-(b). One can determine that when $M_c \parallel M_{FM}$, $V_M < 0$, and when $M_c \parallel -M_{FM}$, $V_M > 0$. Thus, by adjusting the external magnetic field to vary $M_{FM}$, a high voltage when $M_{FM} \parallel -M_c$ and a low voltage when $M_{FM} \parallel M_c$ can be measured by the FM probe.

This approach can be used to detect *not only* the spin magnetization, *but also* the orbital magnetization in WTe₂. As illustrated in Fig. S4(c), the orbital magnetic moment can be converted into spins through the spin-orbit coupling [49]. Therefore, the orbital polarization in WTe₂ can induce spin chemical potential imbalance in the FM probe, leading to a measurable $V_M$ related to the orbital magnetization in WTe₂.

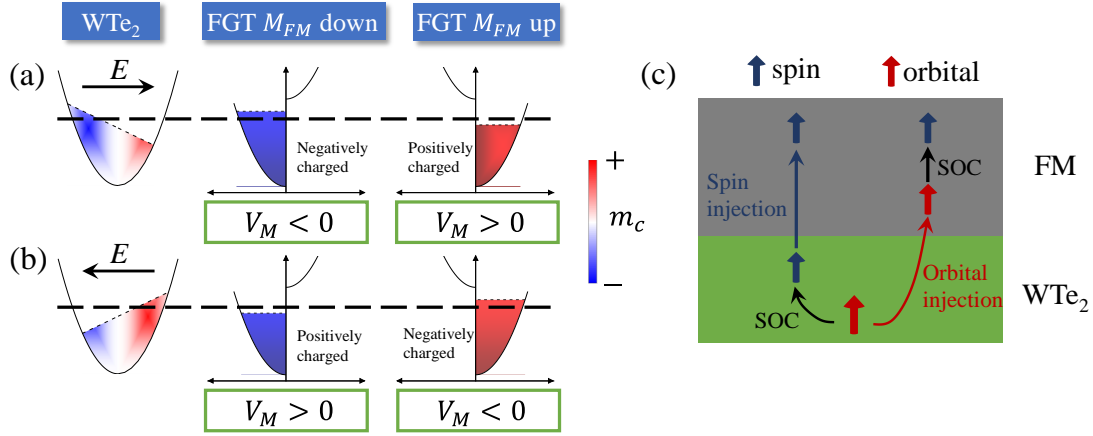

**Fig. S4. Illustration of the principle of magnetization detection using a ferromagnetic probe.**

**a,b,** Left part: Application of an electric field induces polarization of the magnetic moment ($m_c$) in WTe$_2$. Right part: The magnetization of the Fe$_3$GeTe$_2$ (FGT) probe ($M_{FM}$) that can be altered by an external magnetic field. The electrochemical potential balance between states with the same $m_c$ in WTe$_2$ and FGT results in charge transfer, and thus a voltage ($V_M$) can be measured by the FGT probe.

**c, The conversion between orbital (red arrows) and spin (blue arrows) magnetic moments in WTe$_2$/ferromagnet bilayers.** There exist two mechanisms for spin-orbital conversion. For the first channel, the orbital magnetic moment is converted into spin through the spin-orbit coupling (SOC) in WTe$_2$ and then the polarized spin is injected into the FM probe. For the second channel, the orbital angular momentum is injected into FM probe and then is converted into spin through the SOC of the FM probe. For both ways, the orbital magnetization in WTe$_2$ can induce spin chemical potential imbalance in the FM probe.

**The choice of ferromagnetic probe electrode:**

In this work, we opt for a few-layer $Fe_3GeTe_2$ (FGT) flake, a 2D layered van der Waals ferromagnet, as the magnetic probe for measuring the current-induced magnetization in $WTe_2$. The orbital magnetization in $WTe_2$ aligns out-of-plane due to the confinement of charge carriers within a 2D plane. Given the perpendicular magnetic anisotropy of FGT, it is particularly suited for detecting the current induced out-of-plane magnetization. To provide further clarity, Fig. S5 shows the hysteresis curves of the Hall resistance of the FGT as sweeping the in-plane and out-of-plane magnetic fields, respectively. As alterations in the magnetization state of FGT can impact the Hall resistance due to the anomalous Hall effect, the hysteresis curves in Fig. S5 indeed reflects the magnetization variation. The distinct transitions observed in the hysteresis curves only when a magnetic field is applied along the out-of-plane direction confirm the perpendicular magnetic anisotropy of FGT in our device.

Additionally, the Curie temperature of few-layer FGT used in this work is ~180 K [50]. In our study, the temperature for probing current-induced magnetization is below 50 K, ensuring that few-layer FGT meets the requirements for this investigation. During the measurement of current-induced magnetization in $WTe_2$, we introduced a perpendicular magnetic field to manipulate the magnetization state of FGT. This enabled us to detect the magnetization-dependent voltage, representing FGT magnetization in both parallel and antiparallel orientations with respect to the current-induced magnetization. The magnitude of the applied magnetic field exceeded the coercive force of FGT, ensuring that the magnetic domain distribution did not influence the measurement results.

In our experiments, we strategically incorporate a thin *h*-BN layer between the FGT and WTe$_2$ layers. Its role is to act as an insulating barrier, preventing the magnetic proximity effect from FGT on the neighboring WTe$_2$ layer. This ensures that any observed magnetic signals originate solely from current-induced magnetization in WTe$_2$, without influence from FGT's magnetism.

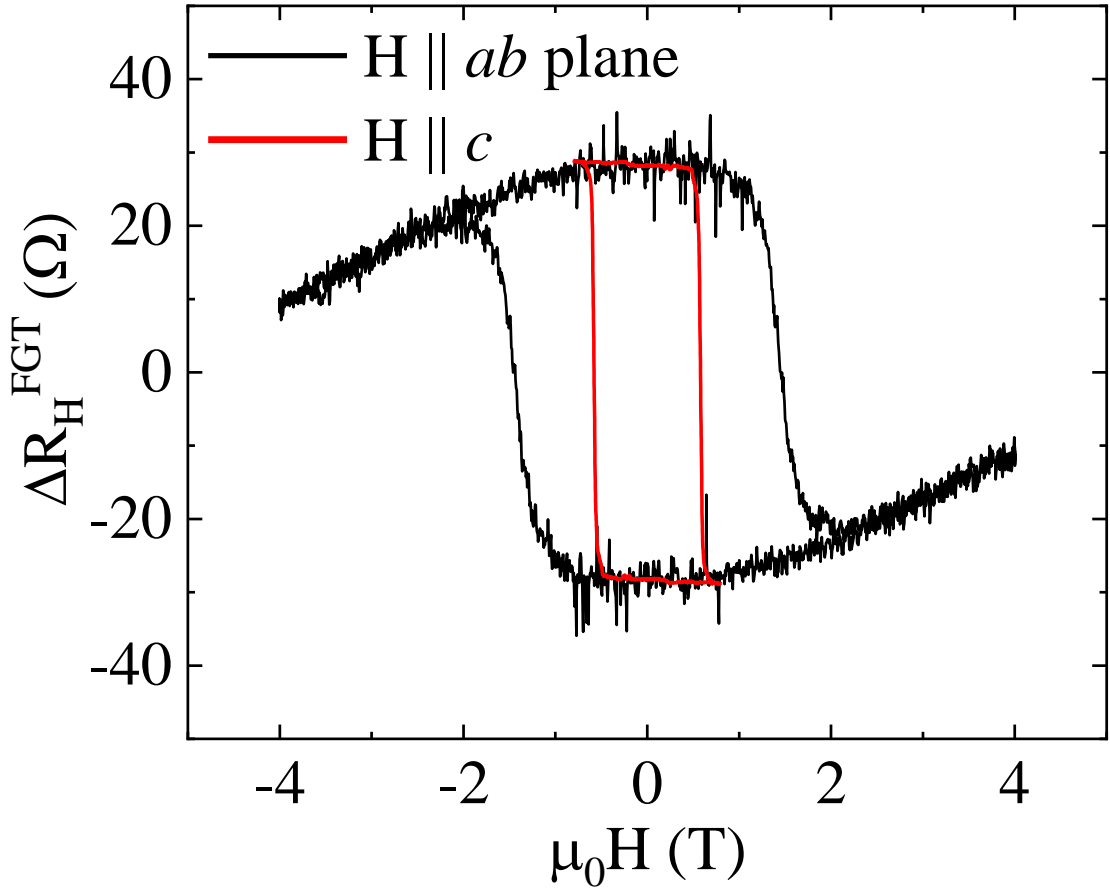

**Fig. S5. The hysteresis curve of the Hall resistance of FGT as sweeping in-plane (black) and out-of-plane (red) magnetic field.**

# Supplemental Note 4: Measurement Scheme for the nonlinear orbital Edelstein effect

The measurement scheme for the magnetization detection is shown in the **Fig. S6**. The disc-shaped electrodes 1-12 are used to apply current along different directions (*i.e.*, along different angle $\varphi$). Electrodes 13 and 14 are voltage contacts. Electrode 13 connects to the $WTe_2$, which is the reference non-magnetic electrode, while the electrode 14 connects to the ferromagnetic probe $Fe_3GeTe_2$, which is regarded as the magnetic probe (see also Fig. S6(b)). **Figure S6** shows the measurement configuration for current along $\varphi = 0°$. To rotate the current direction, we just rotate the current electrodes, for example, from electrodes 1/7 (I+/I-) to electrodes 2/8 (I+/I-), while keep the voltage contacts 13 and 14 unchanged.

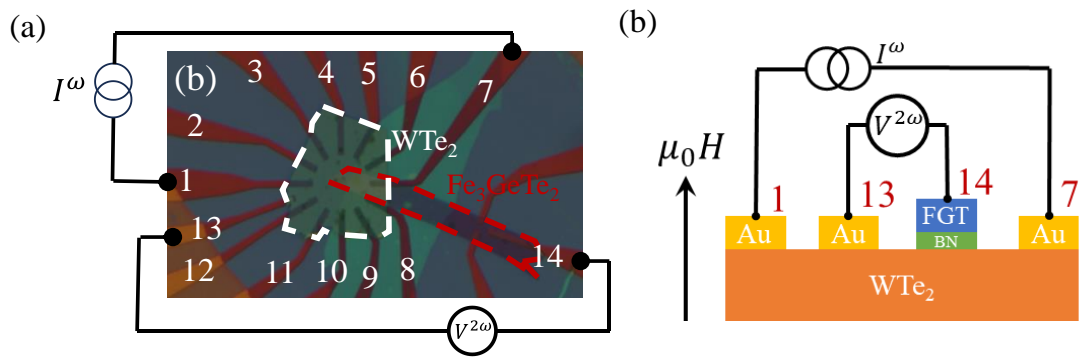

**Fig. S6.** (a) Measurement scheme for the nonlinear orbital Edelstein effect. (b) The illustration of measurement configuration corresponding to (a). The red numbers denote the electrodes depicted in (a).

**Supplemental Note 5: Additional results of device S1**

To investigate current-induced out-of-plane magnetization in WTe$_2$, a magnetization-dependent voltage is collected via a FGT probe. An AC current is injected using the standard lock-in method, and the measured voltage is shown in Fig. S7. Figure S7 shows the current-dependence of the measured second-harmonic voltage $V^{2\omega}$. To obtain the magnetization-related voltage signals $V_m^{2\omega}$, a linear background is removed from $V^{2\omega}$.

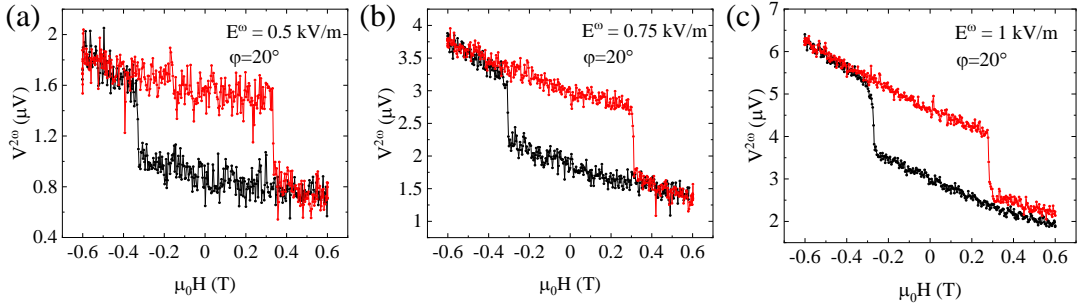

**Fig. S7. The measured second-harmonic voltage ($V^{2\omega}$) via the FGT probe under varying excitation current at fixed direction ($\varphi = 20°$).**

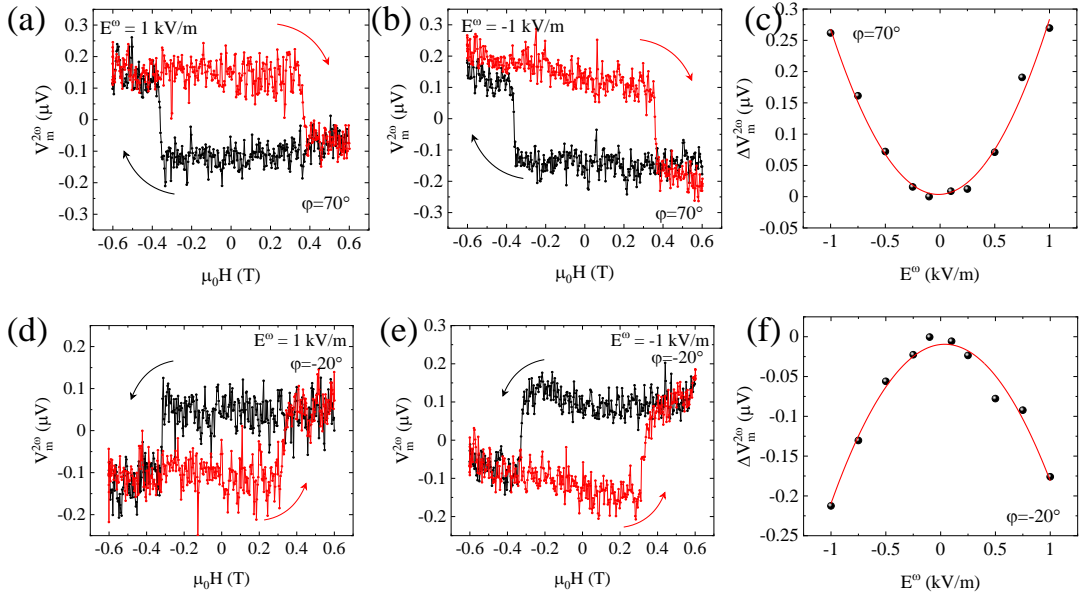

**Fig. S8. Current induced second-harmonic out-of-plane magnetization of WTe$_2$ at 5 K.**

**a,b,** Second-harmonic magnetization-dependent voltage $V_m^{2\omega}$ at $\varphi = 70°$ under (a) $E^\omega = 1$ kV/m and (b) $E^\omega = -1$ kV/m, respectively.

**c,** The loop height of the $V_m^{2\omega}$ - H hysteresis curve, $\Delta V_m^{2\omega}$, as a function of $E^\omega$ at $\varphi = 70°$.

**d,e,** $V_m^{2\omega}$ at $\varphi = -20°$ under (d) $E^\omega = 1$ kV/m and (e) $E^\omega = -1$ kV/m, respectively.

**f,** $\Delta V_m^{2\omega}$ as a function of $E^\omega$ at $\varphi = -20°$.

The second-harmonic magnetization-dependent voltage ($V_m^{2\omega}$) is measured at different angles, all showing consistent results, as shown in Fig. S8. With a fixed source-drain current, we measured the $V_m^{2\omega}$ voltage drop between the FGT electrode and various reference Au electrodes. The results are shown in Fig. S9. It is found that the $V_m^{2\omega}$ shows negligible dependence on the reference voltage probe configurations.

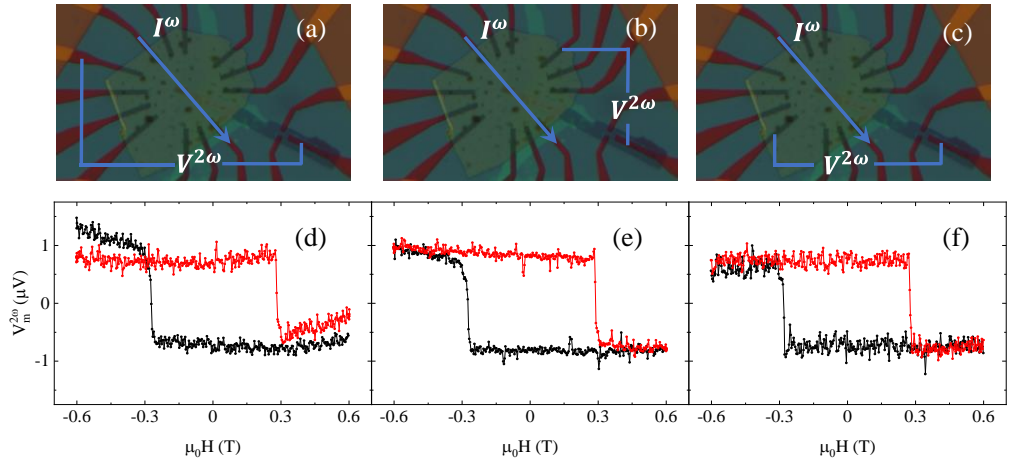

**Fig. S9. Second-harmonic magnetization-dependent voltage $V_m^{2\omega}$ under different reference Au electrodes.**

**a-c,** The voltage measurement configurations with the fixed FGT probe and different reference Au electrodes.

**d-f,** The measured $V_m^{2\omega}$ under corresponding measurement configurations in **a-c**, respectively.

## Supplemental Note 6: Reproducible results in device S2

We fabricated Device S2 with a similar configuration to Device S1. As presented in Fig. S10, Device S2 exhibits consistent results with Device S1 in terms of its second-harmonic voltage, $V_m^{2\omega}$.

The angle dependent measurement was implemented in device S2. The results are presented in Figs. S11. The angle-dependence of resistance and third-order nonlinear Hall effect is well consistent with previous works [44]. The angle-dependence of $R_{xx}$ is fitted by

$$R_{xx}(\theta) = R_a\cos^2(\varphi - \varphi_0) + R_b\sin^2(\varphi - \varphi_0),$$

where $R_a$ and $R_b$ are resistance along crystalline $a$ and $b$ axis, respectively. The third-order nonlinear Hall effect was measured via standard lock-in method. The third-order nonlinear Hall effect shows angle-dependence following the formula

$$\frac{E_H^{3\omega}}{(E^\omega)^3} \propto \frac{\cos(\varphi-\varphi_0)\sin(\varphi-\varphi_0)[(\chi_{22}r^4-3\chi_{12}r^2)\cos^2(\varphi-\varphi_0)+(3\chi_{21}r^2-\chi_{11})\sin^2(\varphi-\varphi_0)]}{(\sin^2(\varphi-\varphi_0)+r\cos^2(\varphi-\varphi_0))^3},$$

where $E_H^{3\omega} = \frac{V_H^{3\omega}}{W}$, $E^\omega = \frac{I^\omega R_{xx}}{L}$, $V_H^{3\omega}$ is the third-harmonic Hall voltage, $I^\omega$ is the applied current, $W$ and $L$ are channel width and length, respectively, $r$ is the resistance anisotropy, $\chi_{ij}$ are elements of the third-order susceptibility tensor, $\varphi_0$ is the angle misalignment between $\varphi = 0°$ and the crystalline $a$ axis. The fitting curves yield a misalignment between the electrode base line and the crystalline axis, where the $a$-axis is at around $\varphi = 6°$ in device S1 and $\varphi = 12°$ in device S2. Importantly, the $\Delta V_m^{2\omega}$ as a function of angle $\varphi$ also shows the same symmetry as the third-order nonlinear Hall effect (Fig. S11(c)), which is consistent with the BCP mechanism in WTe$_2$. In addition to the second-harmonic magnetic voltage $\Delta V_m^{2\omega}$, the first-harmonic magnetic voltage is also observed (Fig. S11(d)), consistent with previous works [54] about the linear Edelstein effect in WTe2. The angle-dependence of the first-harmonic signal shows a sine-shaped curve with maximum value along *a* axis, and zero value along *b* axis. This is distinct from the second-harmonic signal where zero values are observed along both *a* and *b* axis. The different symmetries indicate the different physical origin of the first- and second-harmonic signals.

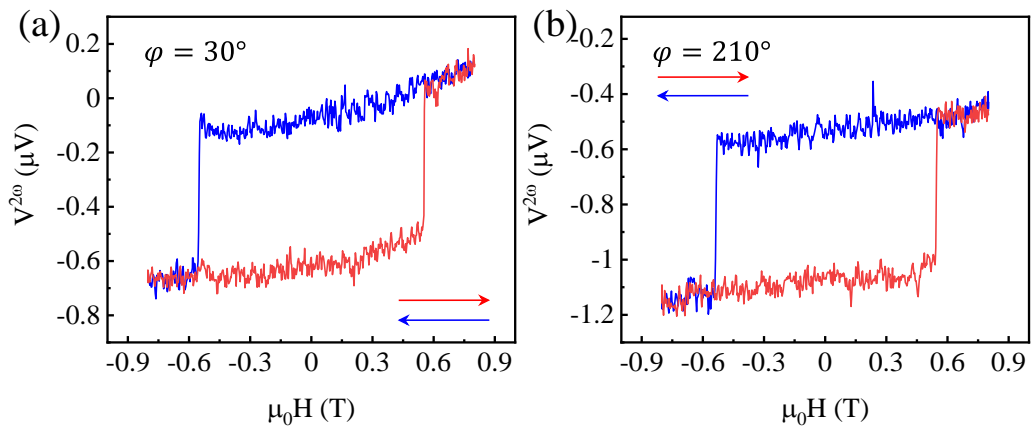

**Fig. S10. The measured second-harmonic voltage ($V^{2\omega}$) via the FGT probe under $E^{\omega} = 1$ kV/m at (a) $\varphi = 30°$ and (b) $\varphi = 210°$ in device S2 at 5 K.**

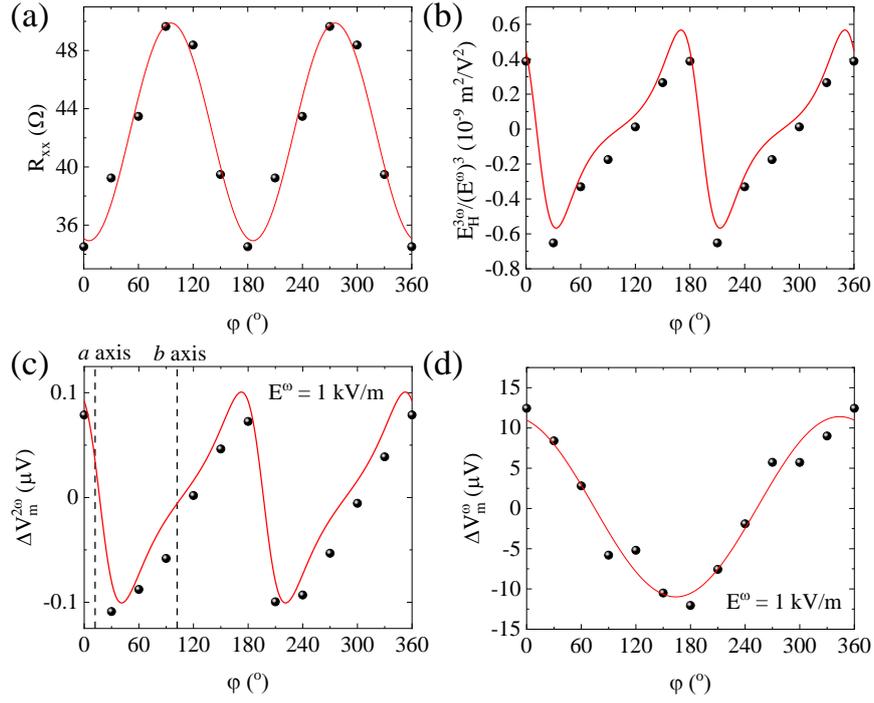

**Fig. S11. Anisotropic transport in device S2.**

**a-d,** (a) Longitudinal resistance $R_{xx}$, (b) third-order nonlinear Hall effect $\frac{E_H^{3\omega}}{(E^\omega)^3}$, (c) second-harmonic magnetization signal $\Delta V_m^{2\omega}$, and (d) first-harmonic magnetization signal $\Delta V_m^\omega$ as a function of angle $\varphi$ at 5 K.

**Supplemental Note 7: Discussions about the origin of the current-induced nonlinear magnetization.**

1. **Contact-induced nonlinearity.**

    The nonlinear voltage can be induced by contact-induced diode effect. However, as shown in Fig. S12(a), the linear I-V relationship is observed, indicating the ohmic nature of the contacts. Therefore, the scenario of contact-induced nonlinearity can be ruled out.

2. **Nonlinear spin/orbital Hall effect**

    For the spin/orbital Hall effect, under the application of a current, the spin/orbit, characterized with *z*-component magnetic moments, deflects towards opposing directions and accumulates respectively on two distinct edges of the sample. If there is a nonlinear version of the spin/orbital Hall effect, the measured $V_m^{2\omega}$ should be sensitive to the configurations of the probe electrodes. This sensitivity arises from the influence of spin accumulation at the reference electrode, impacting the measured signal magnitude. In our experiments, as demonstrated in Fig. S9, the $V_m^{2\omega}$ exhibits negligible dependence on the position of reference electrode. In this setup, the FGT electrode is fixed, while the reference electrodes are positioned on the left and right sides of the source-drain current. Hence, the nonlinear spin/orbital Hall effect is not the origin of the observed phenomenon in our experiment.

3. **Spin Seebeck effect**

    Applying a current in WTe$_2$, a temperature gradient $\nabla T \propto I_e^2$ may be induced. Under this temperature gradient, a spin-polarized current may emerge if there are

differences in the Seebeck coefficients between the spin-up and spin-down states. Previous reports indicate that a current along the *a*-axis can linearly generate the spin polarization along *z*-direction in WTe$_2$ [51-54], implying the presence of a specialized spin texture with some form of spin-momentum locking. Consequently, if the temperature gradient has a component along the *a*-axis, out-of-plane spin polarization may be induced due to the spin Seebeck effect, proportional to $\nabla T \propto I_e^2$.

In our experiments, as illustrated in Fig. S11 for the results of device S2, the $\Delta V_m^{2\omega}$ exhibits the same symmetry as the third-order nonlinear Hall signals $\frac{E_H^{3\omega}}{(E^\omega)^3}$, aligning with the the BCP mechanism in WTe$_2$. Crucially, the third-order nonlinear Hall signal is zero along the *a* or *b* axis due to the symmetry of WTe$_2$ [44]. Similarly, the $\Delta V_m^{2\omega}$ also approaches a minimum along the *a* or *b* axis. This observation contradicts the expectations of the spin Seebeck effect, where the maximum is anticipated along the *a*-axis due to a specific spin texture. Therefore, the spin Seebeck effect can be ruled out as the origin of the observed second-harmonic magnetization signal.

### 4. Nonlinear Orbital and Spin Edelstein effect

In this work, the observed second-harmonic magnetization signals are attributed to the nonlinear orbital and spin Edelstein effect through the BCP tensors. Upon applying a current, the BCP tensor can result in Berry connection linearly proportional to the applied current, that is, $\mathbf{A}^{(1)} = \overleftrightarrow{\mathbf{G}} \mathbf{E}$. Here $\overleftrightarrow{\mathbf{G}}$ is the BCP tensors, $\mathbf{E}$ is the applied electric field, and the superscript "(1)" represents that the physical quantity is the first order term of electric field. Moreover, the field-induced Berry connection would contribution to a correction term of orbital magnetic moment $\boldsymbol{m}_{\boldsymbol{k}}^{(1)} =$ $-\frac{e}{2\hbar}\langle n|(\overleftrightarrow{\boldsymbol{G}}\boldsymbol{E})\times\boldsymbol{\nabla}_{\boldsymbol{k}}H|n\rangle$. The field-induced orbital magnetic moment shows antisymmetric distributions in the momentum space with orbit-momentum locking, thus inducing nonzero magnetoelectric coefficient $\alpha_{cj}^{(1)}=\int_{\boldsymbol{k}}\frac{df}{d\varepsilon}m_{\boldsymbol{k},c}^{(1)}v_{\boldsymbol{k},j}\propto E$. Therefore, orbital magnetization is induced by the electric field $\mathbf{E}$ with $M_c=\alpha_{cj}^{(1)}E_j\propto E^2$, that is, nonlinear orbital Edelstein effect.

In the results of device S2 illustrated in Fig. S11, the symmetry of $\Delta V_m^{2\omega}$ is consistent with that of the third-order nonlinear Hall signals ($\frac{E_H^{3\omega}}{(E^\omega)^3}$) (Fig. S12(b)). Furthermore, a linear relationship exists between $\Delta V_m^{2\omega}$ and the third-order nonlinear Hall effect, as depicted in Fig. 4(c). These findings suggest a common origin for both the current-induced second-harmonic magnetization and the third-order nonlinear Hall effect, namely, the BCP mechanism. Through this BCP mechanism, the current generates the polarization of orbital magnetic moments by modifying Berry connection.

It is noteworthy that, owing to the presence of spin-orbit coupling in WTe$_2$, the orbital magnetization gives rise to a spin polarization, leading to a corresponding nonlinear spin Edelstein effect. While it can be challenging to disentangle the contributions of orbit and spin in systems with strong spin-orbit coupling, the fundamental essence of this observation is rooted in the orbital effect. The spin contribution arises as a consequence of spin-orbit coupling.

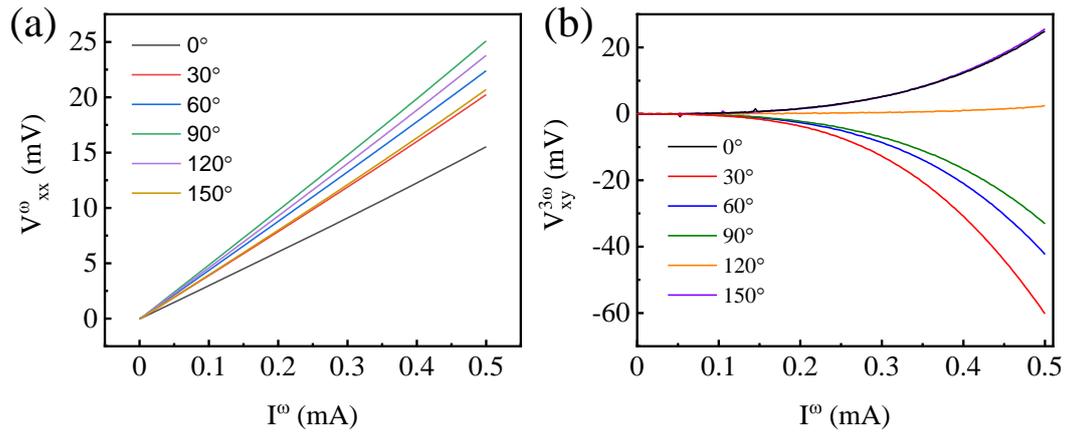

**Fig. S12.** (a) The characteristic IV curves of device S2 at 5 K. (b) Third-order nonlinear Hall effect at 5 K in device S2.